\documentclass[twocolumn]{aastex62}

\usepackage{graphicx}	
\usepackage{amsmath}	
\usepackage{amssymb}	
\usepackage{xcolor}	    

\shorttitle{$R$-matrix collision data for H- and He-like ions}
\shortauthors{Mao et al.}

\begin{document}

\title{$R$-matrix electron-impact excitation data for the H- and He-like ions with $Z=6-30$}

\correspondingauthor{Junjie Mao}
\email{jmao@tsinghua.edu.cn}

\author[0000-0001-7557-9713]{Junjie Mao}
\affiliation{Department of Astronomy, Tsinghua University, Beijing 100084, China}
\affiliation{Department of Physics, Hiroshima University, 1-3-1 Kagamiyama, HigashiHiroshima, Hiroshima 739-8526, Japan}
\affiliation{Department of Physics, University of Strathclyde, Glasgow G4 0NG, UK}

\author[0000-0002-4125-0204]{G. Del Zanna}
\affiliation{Department of Applied Mathematics and Theoretical Physics, University of Cambridge, Cambridge CB3 0WA, UK}

\author[0000-0001-9911-7038]{Liyi Gu}
\affiliation{SRON Netherlands Institute for Space Research, Niels Bohrweg 4, 2333 CA Leiden, the Netherlands}
\affiliation{RIKEN High Energy Astrophysics Laboratory, 2-1 Hirosawa, Wako, Saitama 351-0198, Japan}

\author{C. Y. Zhang}
\affiliation{Department of Physics, University of Strathclyde, Glasgow G4 0NG, UK}

\author{N. R. Badnell}
\affiliation{Department of Physics, University of Strathclyde, Glasgow G4 0NG, UK}







\begin{abstract}
Plasma models built on extensive atomic data are essential to interpreting the observed cosmic spectra. H-like Lyman series and He-like triplets observable in the X-ray band are powerful diagnostic lines to measure the physical properties of various types of astrophysical plasmas. Electron-impact excitation is a fundamental atomic process for the formation of H-like and He-like key diagnostic lines. Electron-impact excitation data adopted by the widely used plasma codes (AtomDB, CHIANTI, and SPEX) do not necessarily agree with each other. Here we present a systematic calculation of electron-impact excitation data of H-like and He-like ions with the atomic number $Z=6-30$ (i.e., C to Zn). Radiation damped $R$-matrix intermediate coupling frame transformation calculation was performed for each ion with configurations up to $n=6$. We compare the present work with the above three plasma codes and literature to assess the quality of the new data, which are relevant for current and future high-resolution X-ray spectrometers.
\end{abstract}


\section{Introduction}
\label{sct:intro}
X-ray emitting hot astrophysical plasmas are ubiquitous in the Universe: stellar coronae, supernova remnants, hot plasmas in individual galaxies and galaxy assemblies, and the warm-hot intergalactic media along the cosmic web filaments \citep{Kaastra2017}. When these targets are observed with spectrometer aboard X-ray space observatories (e.g., Chandra, XMM-Newton, and Suzaku), prominent H- and He-like emission lines from various elements (e.g., O and Fe) often stand out above the continuum \citep[e.g.,][]{Paerels2003, Mao2019c}. These emission lines are powerful diagnostics tools to constrain the physical properties of the hot astrophysical plasmas, such as temperature, density, elemental abundance, and kinematics.

From the observational perspective, we will soon enter an era with the next generation of X-ray spectrometers, including X-ray Imaging and Spectroscopy Mission \citep[XRISM,][to be launched in early 2023]{Tashiro2018}, Advanced Telescope for High Energy Astrophysics \citep[Athena,][to be launched in the 2030s]{Nandra2013,Barret2018},
Arcus \citep[][proposed in the U.S.]{Smith2016}, Hot Universe Baryon Surveyor \citep[HUBS,][proposed in China]{CuiW2020}, Super-Diffuse Intergalactic Oxygen Surveyor \citep[Super-DIOS,][proposed in Japan]{Yamada2018}, Colibr\'{i} \citep[][proposed in Canada]{Heyl2019}, and so on.

We had a taste of the future with the Soft X-ray Spectrometers \citep[SXS, ][]{Mitsuda2014} aboard Hitomi. When observing the hot ($\sim4.6\times10^{7}$~K) intracluster media (ICM) of the Perseus galaxy cluster, dozens of emission lines from various ionization stages of cosmic abundant (e.g., Si, Fe, and Ni) and rare (e.g., Cr and Mn) elements are observed. The high-quality line-rich spectrum was used to study the line-of-sight turbulent velocity dispersion \citep{Hitomi-C2016}; the origin of cosmic elements in the ICM \citep{Hitomi-C2017}; the resonance scattering effect of the ICM \citep{Hitomi-C2018V}; and the temperature structure of the ICM \citep{Hitomi-C2018T}.

Astrophysical plasma models play a vital role in interpreting the observed high-resolution X-ray spectra \citep{Raymond2005,Kaastra2008}. When modeling hot astrophysical plasmas in the collisional ionized equilibrium (CIE), both the APEC \citep{Smith2001,Foster2012} model (and its variants) in XSPEC \citep{Arnaud1996} and the CIE model in SPEX \cite{Kaastra1996,Kaastra2020Z} are widely used in the community. CHIANTI \citep{Dere1997,DelZanna2021} can also model CIE plasma and it is widely used in the solar community. All these plasma models are built on an extensive yet ever-expanding atomic database. High-quality X-ray spectra from future missions are challenging the plasma models developed since the 1970s \citep{Landini1970,Mewe1972,Raymond1977}.

When analyzing the same Hitomi/SXS spectra of Perseus using different plasma models \citep{Hitomi-C2018A}, the measured Fe abundance was found to differ by 16\%. The systematic uncertainty due to the instrumental effects (e.g., effective area uncertainty and gain correction factor) is within 15\%. The statistical uncertainty is, however, about 1\%. That is to say, the power of the instrument is not fully exploited. Theoretical atomic calculations and laboratory measurements of the atomic data \citep[e.g.,][]{BMartinez2019,BMartinez2020,GuL2019,GuL2020,Heuer2021,Shah2021} are required to bring the results of plasma diagnostics closer.

In this work, we focus on the electron-impact excitation (EIE) data for H- and He-like ions from C to Zn. Electron-impact excitation is one of the fundamental atomic processes in astrophysical plasmas. During the collision between a free electron and an ion, energy can be transferred from the free electron to a bounded electron in the ion, exciting it to an upper energy level. When the excited electron decays back to the lower level via radiative transition, at least one photon is emitted and contributes to the emission lines in the observed spectra.

\section{Diagnostic lines and line power}
\label{sct:line_power}
H-like Lyman series and He-like triplets (Table~\ref{tbl:line_list}) are the key diagnostic lines to measure the physical properties of astrophysical plasmas. These lines are in general strong in the observed spectra \citep[cf. the review of solar diagnostics][]{DelZanna2018}. Caution that to properly model the observed spectra, dielectronic satellite lines of He-like lines should be included \citep{Dere2019}, which is beyond the scope of this paper.

\begin{table}
\centering
\footnotesize
\caption{Key diagnostics transitions for H-like and He-like ions.  }
\label{tbl:line_list}
\begin{tabular}{l|ccc} 
\hline
Label & Lower level & Upper level \\
\hline
Ly$\alpha$ (H-like) & ${\rm 1s~^2S_{1/2}}$ & ${\rm 2p~^2P_{3/2,1/2}}$ \\
Ly$\beta$ (H-like) & ${\rm 1s~^2S_{1/2}}$ & ${\rm 3p~^2P_{3/2,1/2}}$ \\
Ly$\gamma$ (H-like) & ${\rm 1s~^2S_{1/2}}$ & ${\rm 4p~^2P_{3/2,1/2}}$ \\
Ly$\delta$ (H-like) & ${\rm 1s~^2S_{1/2}}$ & ${\rm 5p~^2P_{3/2,1/2}}$ \\
He$\alpha$-w (He-like) & ${\rm 1s^2~^1S_0}$ & ${\rm 1s~2p~^1P_1}$ \\
He$\alpha$-x (He-like) & ${\rm 1s^2~^1S_0}$ & ${\rm 1s~2p~^3P_2}$ \\
He$\alpha$-y (He-like) & ${\rm 1s^2~^1S_0}$ & ${\rm 1s~2p~^3P_1}$ \\
He$\alpha$-z (He-like) & ${\rm 1s^2~^1S_0}$ & ${\rm 1s~2s~^3S_1}$ \\
He$\beta$-w (He-like) & ${\rm 1s^2~^1S_0}$ & ${\rm 1s~3p~^1P_1}$ \\
He$\gamma$-w (He-like) & ${\rm 1s^2~^1S_0}$ & ${\rm 1s~4p~^1P_1}$ \\
He$\delta$-w (He-like) & ${\rm 1s^2~^1S_0}$ & ${\rm 1s~5p~^1P_1}$ \\
\noalign{\smallskip}
\hline
\end{tabular}
\end{table}

\begin{figure*}
\centering
\includegraphics[width=\hsize, trim={0.cm 1.0cm 1.0cm 0.cm}, clip]{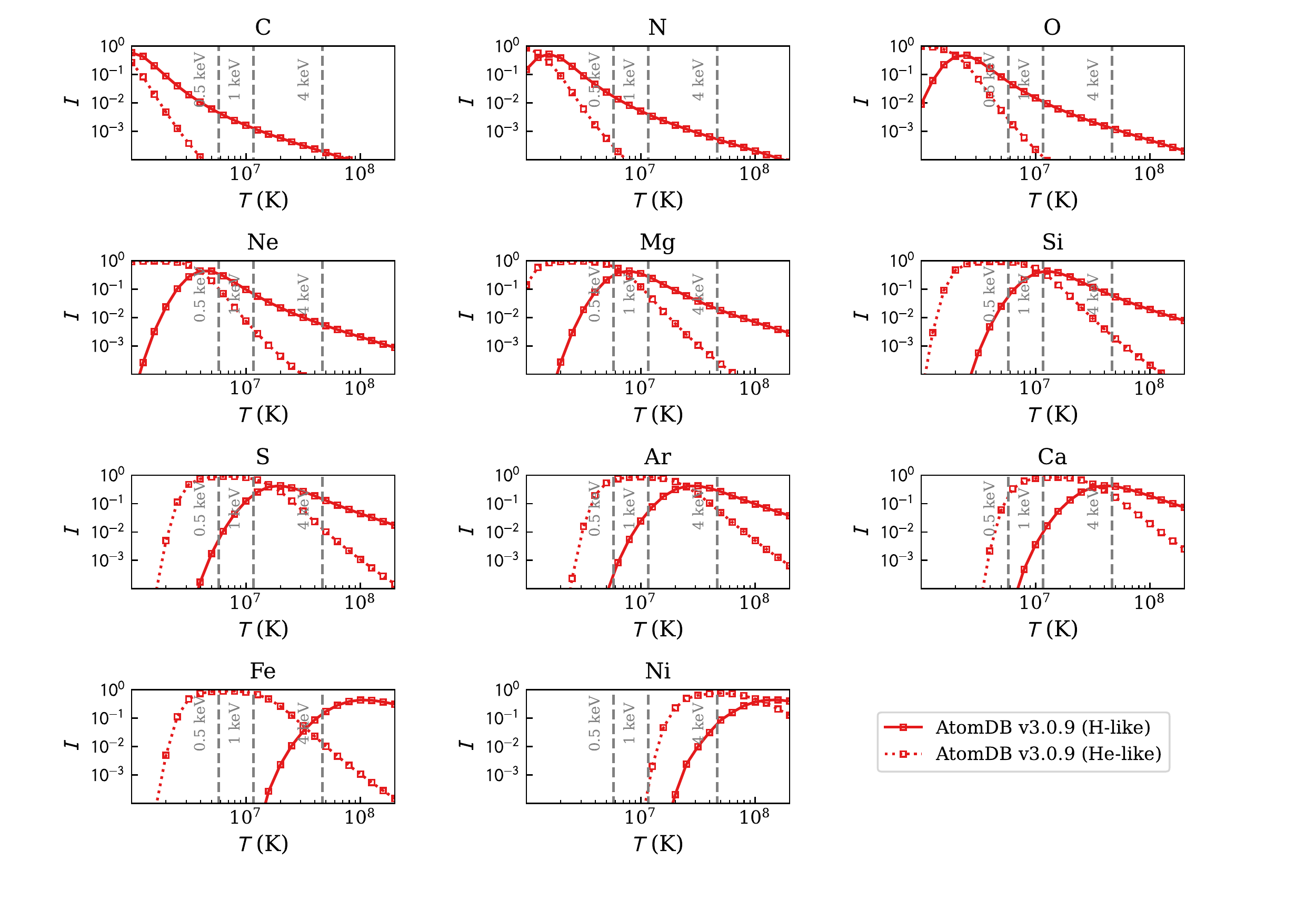}
\caption{Ionic fraction $(I)$ for cosmic abundant metals in CIE plasmas based on \citet{Bryans2009} as the default of AtomDB v3.0.9. The red solid and dashed lines are for H- and He-like ions, respectively. Vertical gray dashed lines marks the typical temperatures of hot plasmas in individual galaxies ($\sim$0.5 keV, dotted), groups of galaxies ($\sim$1 keV, dashed), and clusters of galaxies ($\sim$4 keV, dot-dashed), respectively.}
\label{fig:plot_cfcsd}
\end{figure*}

Lyman series are transitions with $n{\rm p~^2P_{3/2,1/2}} \rightarrow {\rm 1s~^2S_{1/2}}~(n\ge2)$. We mainly focus on Ly$\alpha$ ($n=2\rightarrow1$), Ly$\beta$ ($n=3\rightarrow1$), Ly$\gamma$ ($n=4\rightarrow1$), and Ly$\delta$ ($n=4\rightarrow1$) as they are all available in AtomDB, SPEX, and CHIANTI. For a low-density CIE plasma, the Ly$\alpha$ line (${\rm 2p~^2P_{3/2,1/2}} \rightarrow {\rm 1s~^2S_{1/2}}$) should have the highest line power. However, in a high-density CIE plasma, resonance scattering can reduce the intensity of Ly$\alpha$ by scattering a fraction of photons outside our line-of-sight. This will lead to larger ratios of Ly$\beta$/Ly$\alpha$, Ly$\gamma$/Ly$\alpha$, and Ly$\delta$/Ly$\alpha$ than those in a low-density CIE plasma. On the other hand, at the interface between the hot plasma and cold medium, the charge-exchange process can selectively increase the intensity of e.g., Ly$\gamma$ or Ly$\delta$ \citep{GuL2016}. This also leads to a larger ratio of Ly$\beta$/Ly$\alpha$, Ly$\gamma$/Ly$\alpha$, and Ly$\delta$/Ly$\alpha$ than those in a low-density CIE plasma.

He-like triplet refers to the resonance (allowed) ${\rm 1s~2p~^1P_1} \rightarrow {\rm 1s^2~^1S_0}$, inter-combination (semi-forbidden) ${\rm 1s~2p~^3P_{1,2}} \rightarrow {\rm 1s^2~^1S_0}$, and forbidden ${\rm 1s~2s~^3S_1} \rightarrow{\rm 1s^2~^1S_0}$ transition respectively. The two inter-combination lines are often treated as one line because they are not resolved with current instruments (but will be resolved with future missions). The line ratios among the three are sensitive to plasma temperature and density, external radiation field, and charge-exchange process \citep{Porquet2010}. For a low-density CIE plasma, the resonance line should have the highest line power and the inter-combination line has the lowest line power. In a high-density CIE plasma, on one hand, resonance scattering can reduce the intensity of the resonance line by scattering a fraction of photons outside our line-of-sight \citep[e.g.][]{Sazonov2002, Xu2002, Ogorzalek2017, Hitomi-C2018A, Chen2018}; on the other hand, the inter-combination line will be stronger and the forbidden line will be weaker because collisional excitation will depopulate the upper level of the forbidden line to those of the intercombination lines \citep[e.g.,][]{Porquet2010}. Furthermore, at the interface between the hot plasma and cold medium, the charge-exchange process can increase the forbidden to resonance line ratio \citep[e.g.,][]{BRaymont2007, ZhangSN2014, GuL2016}. Similarly, the line ratio of the He$\alpha$ triplets can also be different from collisionally ionized equilibrium plasma due to photo-excitation \citep[e.g.,][]{Porquet2010}. Higher-order resonance (allowed) ${\rm 1s}~n{\rm p~^1P_1} \rightarrow {\rm 1s^2~^1S_0}$ with $n=3-5$ can also be present in high-quality spectra of future missions, while other higher-order He-like lines are less observable.

For each optically thin emission line in a CIE plasma, its strength can be described by line power $P_{ji}$ (in photons per unit time and volume):
\begin{equation}
    P_{ji}(T,~n_{\rm H}) = A_{ji}~n_{\rm H}~A(Z)~I(T,~n_{\rm H})~N_j(T,~n_{\rm H}),
\end{equation}
where $A_{ji}$ is the spontaneous transition probability from the upper level $j$ to the lower level $i$, $n_{\rm H}$ the hydrogen number density of the plasma, $A(Z)$ the elemental abundance with respect to hydrogen, $Z$ the atomic number, $I$ the normalized ionic fraction (the sum of all the ionization stages of the same element is unity), and $N_j$ the normalized level population of the upper level $j$ (the sum of all the levels is unity). While the $A$-value is independent of the plasma temperature, both the ionic fraction ($I$) and level population ($N_j$) depend on the plasma temperature and density.

Elemental abundances $A(Z)$ are often given in units of solar abundance and there are quite a few of solar abundance tables available to use. Generally speaking, C, N, O, Ne, Mg, Si, S, Ar, Ca, Fe and Ni are the relatively abundant ones for $Z\ge6$ \citep{Anders1989,Asplund2009,Lodders2009}.

Ionic fraction $(I)$ is usually taken from pre-calculated ionization balance tables, which only depend on the temperature of low-density CIE plasma (Fig. \ref{fig:plot_cfcsd}). The default ionization balance is \citet{Bryans2009} for APEC, \citet{Dere2009} for CHIANTI, and \citet{Urdampilleta2017} for SPEX. Around the peak ionic fraction temperatures, the ionic fraction agrees within a few percent among the three codes. At both higher and lower temperature ends when the ionic fraction is rather small, larger deviations ($\gtrsim20$\%) can be found. We caution that metastable levels will start to be populated as the plasma density increases, which can modify the ionization balance significantly \citep{Dufresne2019, Dufresne2020, Dufresne2021b}.

The user can choose which ionization balance and solar abundance table to use in pyatomdb\footnote{https://github.com/AtomDB/pyatomdb}, ChiantiPy\footnote{https://github.com/chianti-atomic/ChiantiPy}, and SPEX. When comparing plasma models, it is better to use the same solar abundance and ionization balance tables. 

\begin{figure}
\centering
\includegraphics[width=\hsize, trim={1.5cm 0.cm 1.0cm 0.cm}, clip]{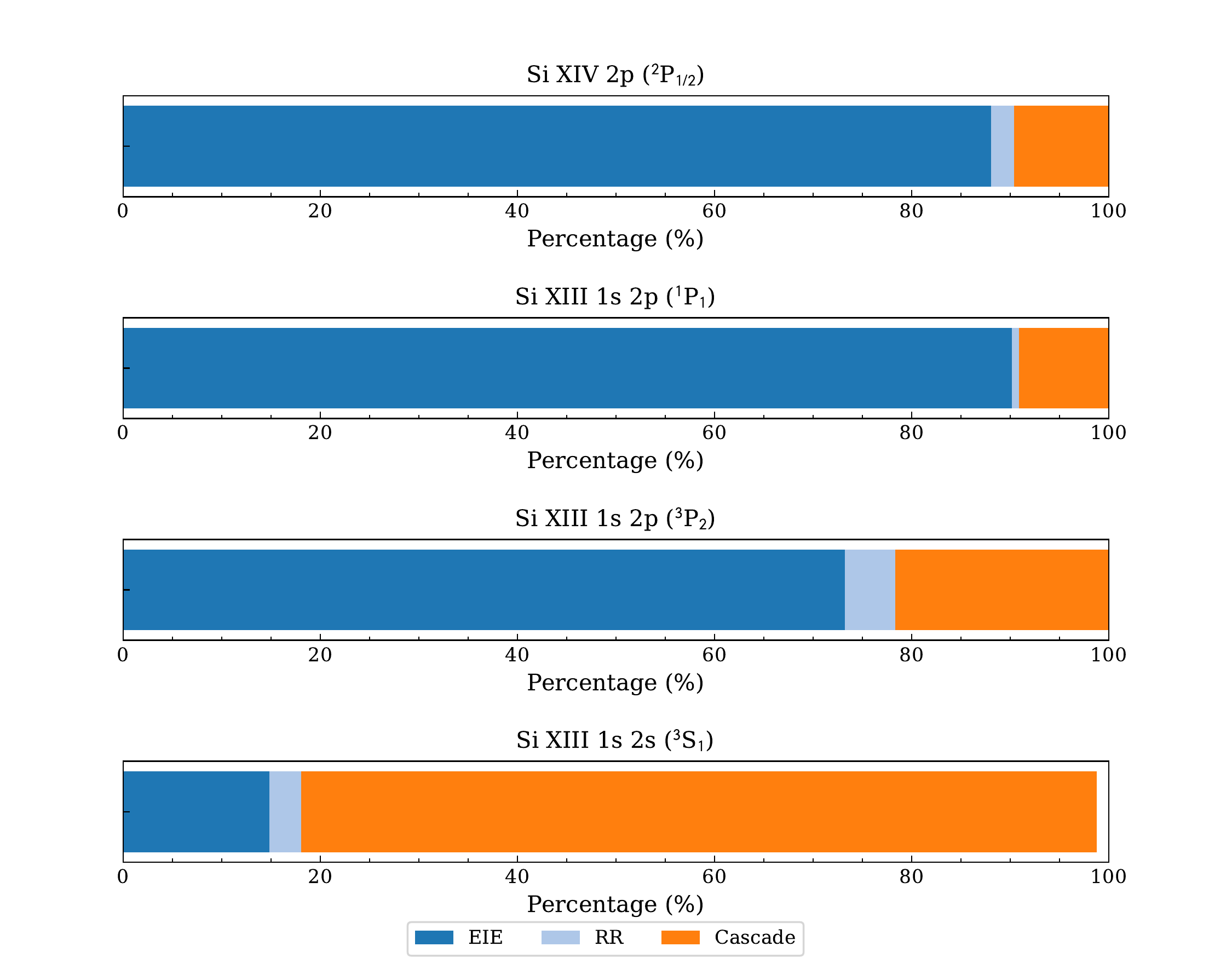}
\caption{Percentage contribution of the atomic processes to the level population of Si {\sc xiv} and Si {\sc xiii} in a CIE plasma with $kT=1$~keV. The four levels, from top to bottom, are the upper levels of Ly$\alpha$, He$\alpha$-w, He$\alpha$-x, and He$\alpha$-z. Contributions from the electron-impact excitation (EIE), radiative recombination (RR), and cascade are shown in dark blue, light blue, and orange. The SPEX code (v3.06.01) is used to calculate the level population here. }
\label{fig:plot_cflpop_z14}
\end{figure}

The level population ($N_j$) depends on various atomic processes. Figure~\ref{fig:plot_cflpop_z14} illustrates the percentage contribution of the atomic processes to the level population of Si {\sc xiv} and Si {\sc xiii} for a CIE plasma with $kT=1$~keV. Similar results can be found for other H- and He-like ions in CIE plasmas. Generally speaking, electron-impact excitation (EIE) contributes most to the upper level population of resonance lines. Radiative recombination (RR) has a minor contribution to the level population. Note that the same RR data, sourced from \citet{Badnell2006}, are implemented via interpolation for AtomDB and CHIANTI or parameterization \citep{Mao2016} for SPEX. The contribution from cascade is negligible for resonance lines but it can be crucial for forbidden lines \citep{Hitomi-C2018A}.

\section{Status quo}
\label{sct:status_quo}
We examine electron-impact excitation data in the latest versions of AtomDB (v3.0.9), CHIANTI (v10.0.1), and SPEX (v3.06.01). The electron-impact excitation data of the key diagnostics line (Table~\ref{tbl:line_list}) are sourced differently in the three atomic databases. For H-like ions, AtomDB mainly adopts the distorted waves data (with the independent process and isolated resonances approximation) of \citet{LiS2015} for elements heavier and including Al. For lighter elements, either $R$-matrix data \citep{Ballance2003} or distorted wave calculation by A. Foster with the Flexible Atomic Code \citep[FAC,][]{GuM2008} are used. For CHIANTI and SPEX, $R$-matrix data are used for a few ions. Interpolation or extrapolation along the iso-electronic sequence is used for the rest of the H-like ions. Table~\ref{tbl:cfsrc_hlike} provides a summary of the source of the electron-impact excitation data of the Ly$\alpha$ to Ly$\delta$ transitions in the three atomic databases.

\begin{table*}
\centering
\footnotesize
\caption{Source of the electron-impact excitation data of the H-like Ly$\alpha$ to Ly$\delta$ transitions in SPEX v3.06.01, AtomDB v3.0.9, and CHIANTI database v10.0.1. DW and RM are short for distorted wave and $R$-matrix calculations, respectively. }
\label{tbl:cfsrc_hlike}
\begin{tabular}{l|ccc} 
\hline
Ion & SPEX & AtomDB & CHIANTI \\
\hline
C {\sc vi} & \citet[][RM]{Aggarwal1991-C} & \citet[][RM]{Ballance2003} & \citet[][RM]{Ballance2003} \\
\noalign{\smallskip}
N {\sc vii} & Interpolation & FAC (DW) & Interpolation \\
\noalign{\smallskip}
O {\sc viii} & Interpolation & \citet[][RM]{Ballance2003} & \citet[][RM]{Ballance2003} \\
\noalign{\smallskip}
Ne {\sc x} & \citet[][RM]{Aggarwal1991-Ne}, DW for Ly$\beta$ & \citet[][RM]{Ballance2003} & \citet[][RM]{Ballance2003} \\
\noalign{\smallskip}
Na {\sc xi} & Interpolation & FAC (DW) & Interpolation \\
\noalign{\smallskip}
Mg {\sc xii} & Interpolation & FAC (DW) & Interpolation \\
\noalign{\smallskip}
Al {\sc xiii} & Interpolation & \citet[][DW]{LiS2015} & Interpolation \\
\noalign{\smallskip}
Si {\sc xiv} & \citet[][RM]{Aggarwal1992-Si} & \citet[][DW]{LiS2015} & \citet[][RM]{Aggarwal1992-Si} \\
\noalign{\smallskip}
P {\sc xv} & Interpolation & \citet[][DW]{LiS2015} & Interpolation \\
\noalign{\smallskip}
S {\sc xvi} & Interpolation & \citet[][DW]{LiS2015} & Interpolation \\
\noalign{\smallskip}
Cl {\sc xvii} & Interpolation & \citet[][DW]{LiS2015} & Interpolation \\
\noalign{\smallskip}
Ar {\sc xviii} & Interpolation & \citet[][DW]{LiS2015} & Interpolation \\
\noalign{\smallskip}
K {\sc xix} & Interpolation & \citet[][DW]{LiS2015} & Interpolation \\
\noalign{\smallskip}
Ca {\sc xx} & \citet[][RM]{Aggarwal1992-Ca} & \citet[][DW]{LiS2015} & \citet[][RM]{Aggarwal1992-Ca} \\
\noalign{\smallskip}
Cr {\sc xxiv} & Interpolation & \citet[][DW]{LiS2015} & -- -- \\
\noalign{\smallskip}
Mn {\sc xxv} & Interpolation & \citet[][DW]{LiS2015} & -- -- \\
\noalign{\smallskip}
Fe {\sc xxvi} & \citet[][RM]{Kisielius1996}, interpolation for Ly${\delta}$ & \citet[][DW]{LiS2015} & \citet[][RM]{Ballance2002} \\
\noalign{\smallskip}
Ni {\sc xxviii} & Extrapolation & \citet[][DW]{LiS2015} & Extrapolation \\
\noalign{\smallskip}
\hline
\end{tabular}
\end{table*}

For He-like ions, AtomDB mainly adopts $R$-matrix data (including the radiation damping effect) for all the levels up to $n=5$: \citet{Whiteford2001} for He-like Fe {\sc xxv} and \citet{Whiteford2005} for other He-like ions. The latter ones, available on OPEN-ADAS\footnote{https://open.adas.ac.uk/}, were calculated following \citet[][for He-like Ar and Fe only]{Whiteford2001} with some modifications (given in the comment section of the data files). These data are not validated (e.g., comparing to previous calculations) in a peer-reviewed journal publication as the lead author left the field before finishing the project. In particular, He-like Fe {\sc xxv} data of \citet{Whiteford2005} is not consistent with that of \citet{Whiteford2001}. This is described in Sect.~\ref{sct:res} later. CHIANTI also uses a large fraction of these data \citep{Whiteford2005}. But it uses the $R$-matrix data (without the radiation damping effect) of \citet{Aggarwal2009} for Na {\sc x} and interpolation along the iso-electronic sequence for P {\sc xiv} and K {\sc xviii}. SPEX adopts the Coulomb-Born-Exchange data of \citet{Sampson1983}, which ignored resonances.

\begin{table*}
\centering
\footnotesize
\caption{Source of the electron-impact excitation data of the He-like triplet transitions in SPEX v3.06.01, AtomDB v3.0.9, and CHIANTI database v9.0.1. \citet{Sampson1983} and \citet{Whiteford2005} used Coulomb-Born-Exchange and $R$-matrix methods, respectively.}
\label{tbl:cfsrc_helike}
\begin{tabular}{l|ccc} 
\hline
Ion & SPEX & AtomDB & CHIANTI \\
\hline
C {\sc v} & \citet[][CBE]{Sampson1983} & \citet[][RM]{Whiteford2005} & Interpolation \\
\noalign{\smallskip}
N {\sc vi} & \citet[][CBE]{Sampson1983} & \citet[][RM]{Whiteford2005} & Interpolation \\
\noalign{\smallskip}
O {\sc vii} & \citet[][CBE]{Sampson1983} & \citet[][RM]{Whiteford2005} & \citet[][RM]{Whiteford2005} \\ 
\noalign{\smallskip}
Ne {\sc ix} & \citet[][CBE]{Sampson1983} & \citet[][RM]{Whiteford2005} & \citet[][RM]{Whiteford2005} \\
\noalign{\smallskip}
Na {\sc x} & \citet[][CBE]{Sampson1983} & \citet[][RM]{Whiteford2005} & \citet{Aggarwal2009} \\
\noalign{\smallskip}
Mg {\sc xi} & \citet[][CBE]{Sampson1983} & \citet[][RM]{Whiteford2005} & \citet[][RM]{Whiteford2005} \\
\noalign{\smallskip}
Al {\sc xii} & \citet[][CBE]{Sampson1983} & \citet[][RM]{Whiteford2005} & \citet[][RM]{Whiteford2005} \\
\noalign{\smallskip}
Si {\sc xiii} & \citet[][CBE]{Sampson1983} & \citet[][RM]{Whiteford2005} & \citet[][RM]{Whiteford2005} \\
\noalign{\smallskip}
P {\sc xiv} & \citet[][CBE]{Sampson1983} & \citet[][RM]{Whiteford2005} & Interpolation \\
\noalign{\smallskip}
S {\sc xv} & \citet[][CBE]{Sampson1983} & \citet[][RM]{Whiteford2005} & \citet[][RM]{Whiteford2005}  \\
\noalign{\smallskip}
Cl {\sc xvi} & \citet[][CBE]{Sampson1983} & \citet[][RM]{Whiteford2005} & Interpolation \\
\noalign{\smallskip}
Ar {\sc xvii} & \citet[][CBE]{Sampson1983} & \citet[][RM]{Whiteford2005} & \citet[][RM]{Whiteford2005} \\
\noalign{\smallskip}
K {\sc xviii} & \citet[][CBE]{Sampson1983} & \citet[][RM]{Whiteford2005} & Interpolation  \\
\noalign{\smallskip}
Ca {\sc xix} & \citet[][CBE]{Sampson1983} & \citet[][RM]{Whiteford2005} & \citet[][RM]{Whiteford2005}  \\
\noalign{\smallskip}
Cr {\sc xxiii} & \citet[][CBE]{Sampson1983} & \citet[][RM]{Whiteford2005} & \citet[][RM]{Whiteford2005} \\
\noalign{\smallskip}
Mn {\sc xxiv} & \citet[][CBE]{Sampson1983} & \citet[][RM]{Whiteford2005} & -- -- \\
\noalign{\smallskip}
Fe {\sc xxv} & \citet[][CBE]{Sampson1983} & \citet[][RM]{Whiteford2001} & \citet[][RM]{Whiteford2005} \\
\noalign{\smallskip}
Ni {\sc xxvii} & \citet[][CBE]{Sampson1983} & \citet[][RM]{Whiteford2005} & \citet[][RM]{Whiteford2005} \\
\noalign{\smallskip}
\hline
\end{tabular}
\end{table*}

Electron-impact excitation data are usually provided in the form of dimensionless effective collisional strength $\Upsilon_{ij}$). This is obtained by convolving the ordinary collision strength ($\Omega_{ij}$) with the Maxwellian distribution:
\begin{equation}
    \Upsilon_{ij} = \int_0^{\infty} \Omega_{ij}~\exp\left(-\frac{E_f}{kT}\right)~d\left(\frac{E_f}{kT}\right)~,
\label{eq:upsilon}
\end{equation}
where $E_f$ is the scattered electron energy, $k$ the Boltzmann constant, and $T$ the electron temperature of the plasma. Effective collisional strength are usually tabulated on a narrow or wide temperature grid, depending on the original calculations. Interpolation among these temperatures and extrapolation beyond the temperature range are implemented by AtomDB and CHIANTI. For SPEX, the collision data as a function of temperature are implemented via parameterization to cover a wide temperature range \citep{Kaastra2008}.

We caution that the energy levels and spontaneous transition rate (i.e., $A$-values) among these three atomic databases do not necessarily agree. Detailed comparisons are given in Appendix~\ref{sct:cflev_cftran}.

\section{$R$-matrix calculation}
\label{sct:rm_mo}
Here we present a systematic $R$-matrix calculation for H- and He-like ions. $R$-matrix intermediate coupling frame transformation calculation \citep[ICFT][]{Griffin1998} including the effect of radiation damping \citep{Robicheaux1995, Gorczyca1996} was performed for each ion with configurations up to $n=6$. That is to say, 36 levels for H-like ions and 71 levels for He-like ions.

We used the AUTOSTRUCTURE code \citep{Badnell2011} to calculate the target atomic structure. Wave functions were obtained by diagonalizing the Breit-Pauli Hamiltonian \citep{Eissner1974}. We include one-body relativistic terms (mass-velocity, nuclear plus Blume \& Watson spin-orbit, and Darwin) perturbatively. The Thomas-Fermi-Dirac-Amaldi model was used for the electronic potential with $nl$-dependent scaling parameters \citep{Nussbaumer1978}. We set the $nl$-dependent scaling parameters to unity, following \citet{Ballance2002} and \citet{Malespin2011}.


For the scattering calculation, we used radiation damped $R$-matrix ICFT method. We used 110 continuum basis orbitals for H- and He-like ions with configurations up to $n=6$ to cover the energy range from the ground state to $\gtrsim6~I_p$, where $I_p$ is the ionization threshold. This ensures the cross section is close to the asymptotic limit before extrapolating to the infinite limit point.

Angular momenta up to $2J=26$ and $2J=96$ were included for the exchange and non-exchange calculations, respectively. Higher angular momenta (up to infinity) were included following the top-up formula of the Burgess sum rule \citep{Burgess1974} for dipole-allowed transitions and a geometric series for the non-dipole-allowed transitions \citep{Badnell2001a}.

The outer-region exchange calculation of the resonance region used a rather fine energy mesh with the number of sampling points ranges from $\sim1.0\times10^5$ for H-like C {\sc vi} to $\sim5.8\times10^5$ for H-like Zn {\sc xxx} and from $\sim0.8\times10^5$ for He-like C {\sc v} to $5.6\times10^5$ for He-like Zn {\sc xxxix}. Beyond the resonance regions (up to six times the ionization potential), the outer-region exchange calculations were performed with a coarse energy mesh with $\sim 2500$ sampling points. A similar coarse energy mesh was also used for the outer-region non-exchange calculations.

To complete the Maxwellian convolution (Eq.~\ref{eq:upsilon}) at high temperatures, we calculated the infinite-energy Born and dipole line strength limits using AUTOSTRUCTURE. Between the last calculated energy point and the two limits, interpolation was used according to the type of transition in the Burgess--Tully scaled domain \citep[i.e. the quadrature of the reduced collision strength over reduced energy; see][]{Burgess1992}.

\section{Results}
\label{sct:res}
We have obtained radiation damped $R$-matrix electron-impact excitation data for the H- and He-like iso-electronic sequence with $Z=6-30$, where $Z$ is the atomic number, e.g., $Z=14$ for silicon. Our effective collision strengths cover four orders of magnitude in temperature $(z+1)^2(2\times10^2,~2\times10^6)~{\rm K}$, where $z$ is the ionic charge (e.g., $z=10$ for He-like Mg {\sc xi}).

Effective collision strength data are archived according to the Atomic Data and Analysis Structure (ADAS) data class {\it adf04} and are available on Zenodo \href{https://zenodo.org/record/7226828#.Y1CbVexByIU}{DOI: 10.5281/zenodo.7226828}. Optimal interval-averaged ordinary collision strength data are also provided, which can be used for convolution with non-Maxwellian distributions. The ordinary collision strength data files are produced with the latest version of adasexj\footnote{http://www.apap-network.org/codes/serial/misc/adasexj.f}. For each transition, the number of bins (or intervals) is around 100, depending on the width of the resonance region. Moreover, the Zenodo package also includes the input files of the $R$-matrix calculations, binned ordinary collision strength data (in the {\it adf04} format), atomic data and python scripts used to create the figures presented in this manuscript. These data will be used to improve the atomic databases of astrophysical plasma codes, such as AtomDB \citep{Smith2001, Foster2012}, CHIANTI \citep{Dere1997,DelZanna2021}, and SPEX \citep{Kaastra1996,Kaastra2020Z}.

\section{Discussion}
\label{sct:dis}
A scattering calculation using the $R$-matrix ICFT method (Sect.~\ref{sct:rm_mo})
necessarily uses the Breit--Pauli R-matrix structure code. This includes only one-body relativistic operators (excluding QED)\footnote{Breit and QED interactions are absent also from the structure used by the Dirac R-matrix code.}. In addition, it requires the user to supply a unique set of non-relativistic orthogonal radial orbitals from an external atomic structure code. Our atomic structure calculated with AUTOSTRUCTURE for subsequent $R$-matrix scattering calculations is denoted as AS-RM. When compared with other structure calculations (including AUTOSTRUCTURE) which make use of two-body relativistic operators and/or QED and/or non-unique and/or non-orthogonal relativistic orbitals, the AS-RM level energies and A-values are less accurate. For the upper levels of the key diagnostic transitions (Table~\ref{tbl:line_list}), the AS-RM level energies can differ up to $\sim0.05$~\% for H-like and $\sim0.23$~\% for He-like when compared to the three atomic databases (Appendix~\ref{sct:cflev_cftran}). Similarly, by $n=5$ the AS-RM A-values can differ by up to $\sim25$~\% for H-like and $\sim40$~\% for He-like while the A-values of the key transitions among the three databases differ by up to $\sim5$~\% for H-like and $\sim40$~\% (Appendix~\ref{sct:cflev_cftran}). More accurate level energies and A-values than the AS-RM ones can be obtained from AUTOSTRUCTURE as described in Appendix~\ref{sct:cflev_cftran} and we denote them AS-REL. Other sources include \citet[][]{Aggarwal2009,Aggarwal2010a,Aggarwal2010b,Aggarwal2012a,Aggarwal2012b,Aggarwal2013,Malespin2011}.

In this section, we compare the effective collisional strength of the key diagnostic lines in Table~\ref{tbl:line_list} among the present work, all three atomic databases (AtomDB, CHIANTI, and SPEX), and some reference results not incorporated in the three atomic databases. We focus on the following representative elements: Fe (Sect.~\ref{sct:cfecs_z26}), Ca (Sect.~\ref{sct:cfecs_z20}), Si (Sect.~\ref{sct:cfecs_z14}), and O (Sect.~\ref{sct:cfecs_z08}). We also show exemplary impacts on observations (Sect.~\ref{sct:simu}).


\subsection{Fe {\sc xxvi} and Fe {\sc xxv}}
\label{sct:cfecs_z26}
\begin{figure*}
\centering
\includegraphics[width=\hsize, trim={0.5cm 2.cm 0.5cm 0.cm}, clip]{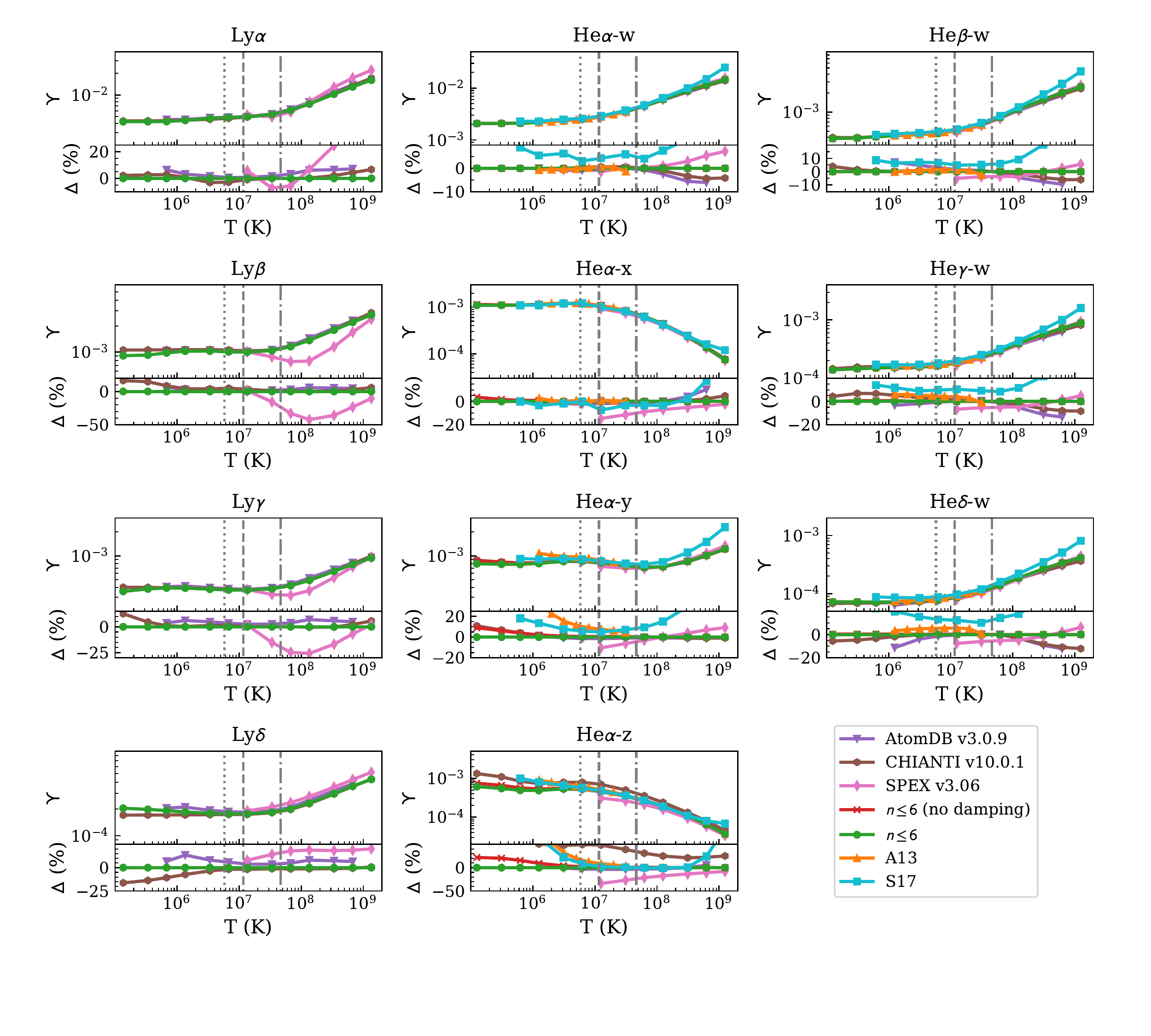}
\caption{Comparison of effective collisional strength data of key diagnostic lines for Fe {\sc xxvi} and Fe {\sc xxv}. AtomDB, CHIANTI, and SPEX data are shown in purple (triangle down), brown (triangle down), and magenta (triangle down), respectively. Present work without radiation damping, present work with radiation damping, \citet{Aggarwal2013}, and \citet{SiR2017} are shown in red (x)}, green (circle), orange (triangle up), and cyan (square), respectively. Percentage difference ($\Delta$) is given with respect to the present work. Vertical dashed lines mark typical temperatures of hot plasmas in individual galaxies ($\sim0.5$~keV, dotted), groups of galaxies ($\sim1$~keV, dashed), and clusters of galaxies ($\sim4$~keV, dot-dashed), respectively. When the ionic fraction is too low (Fig.~\ref{fig:plot_cfcsd}), SPEX skipped the level population calculation (including the effective collisional strength data) for computational efficiency.
\label{fig:plot_cfecs_z26}
\end{figure*}

As shown in Figure~\ref{fig:plot_cfecs_z26}, the effective collision strength of the Lyman series agrees $\lesssim5$~\% for Ly$\alpha$ and $\lesssim20$~\% for Ly$\beta$ to Ly$\delta$ among the present work, AtomDB, CHIANTI, and \citet{Aggarwal2013}. Ly$\alpha$ to Ly$\gamma$ data in SPEX can differ by $\gtrsim50$~\% from the other data sets at $T>10^{7}$~K. The original Dirac $R$-matrix calculation by \citet{Kisielius1996} was performed at $T=10^{6-7.5}$~K. Hence, the root of the difference is in the extrapolation at $T>10^{7.5}$~K in SPEX. Ly$\delta$ in SPEX is obtained from interpolation along the isoelectronic sequence (Table~\ref{tbl:cfsrc_helike}), which is systematically higher (up to $\sim20$~\%) than the present work.

For the He$\alpha$-w (resonance) line, all $R$-matrix data agree $\lesssim3$~\% at $T<10^{8}$~K. Distorted wave data (with independent process and isolated resonances approximation, IPIRDW) from \citet{SiR2017} is systematically higher by a few percent. Such offset is also shown in the Figure 2 of \citet{SiR2017}, where the authors calculated both IPIRDW and Dirac $R$-matrix data. The offset is due to the different treatment of resonances by the two calculations, which is illustrated in Figure 1 of \citet{SiR2017}. At $T>10^8$~K, IPIRDW data by \citet{SiR2017} increases more rapidly than the $R$-matrix data. The difference originates from the convolution of Maxwellian (Eq.~\ref{eq:upsilon}) at high temperatures \citep[cf. Sect.~\ref{sct:rm_mo} and][]{SiR2017}. The comparison of He$\beta$ to He$\delta$ resonance transitions among different data sets share similar issues found for He$\alpha$.

For the He$\alpha$-x (intercombination) line, all $R$-matrix data agree $\lesssim3$~\% at $T<10^{8}$~K. IPIRDW data from \citet{SiR2017} agrees $\lesssim8$~\%. For the He$\alpha$-y (intercombination) line, relatively large differences ($\lesssim25$~\%) can be found among different data sets at $T<10^7$~K. The present work and \citet[][used by AtomDB]{Whiteford2001} agree $\lesssim3$~\% at $T\sim10^{6-7}$~K (the latter does not calculate below $\sim10^{6}$~K). \citet{Whiteford2005} covers one order of magnitude lower in temperature than \citet{Whiteford2001} but differs by up to $\sim15$~\%. Between \citet{SiR2017} and \citet{Aggarwal2013}, the former is relatively lower \citep[see also Figure 2 of][]{SiR2017}. At $T=5.8\times10^6$~K (or 0.5~keV), \citet{Aggarwal2013} is larger by $\sim12$~\% than the present work.

Similarly, large differences are found below $5.8\times10^6$~K for the He$\alpha$-z (forbidden) line. Furthermore, \citet{Whiteford2005} data is systematically above ($\gtrsim20$~\%) all other calculations at $T\lesssim10^8$~K. At $T\sim10^6$~K, \citet{Whiteford2005} is larger than both \citet{Whiteford2001} and the present work by $\sim50$~\%. Such a large difference cannot be explained by the radiation damping effect, which is $\lesssim20$~\% at low temperatures and has no impact at higher temperatures (Figure~\ref{fig:plot_cfecs_z26}). 

For both He$\alpha$-y and z lines, \citet{Aggarwal2013} and \citet{SiR2017} data are larger ($\gtrsim10$~\%) than the present work at $T\lesssim5.8\times10^6$~K (or T$\lesssim0.5$~keV). On the one hand, radiation damping is not included in \citet{Aggarwal2013}. On the other hand, all the calculations are subject to the inherent lack of convergence in the target configuration-interaction expansion and/or the collisional close-coupling expansion for weaker transitions  \citep{FMenchero2017, DelZanna2019}. Note that, under CIE conditions, the ionic fraction of Fe {\sc xxv} at $T\lesssim0.5$~keV is more than three orders of magnitude lower than the peak value (Figure~\ref{fig:plot_cfcsd}).

\subsection{Ca {\sc xx} and Ca {\sc xix}}
\label{sct:cfecs_z20}
\begin{figure*}
\centering
\includegraphics[width=\hsize, trim={0.5cm 2.0cm 0.5cm 0.cm}, clip]{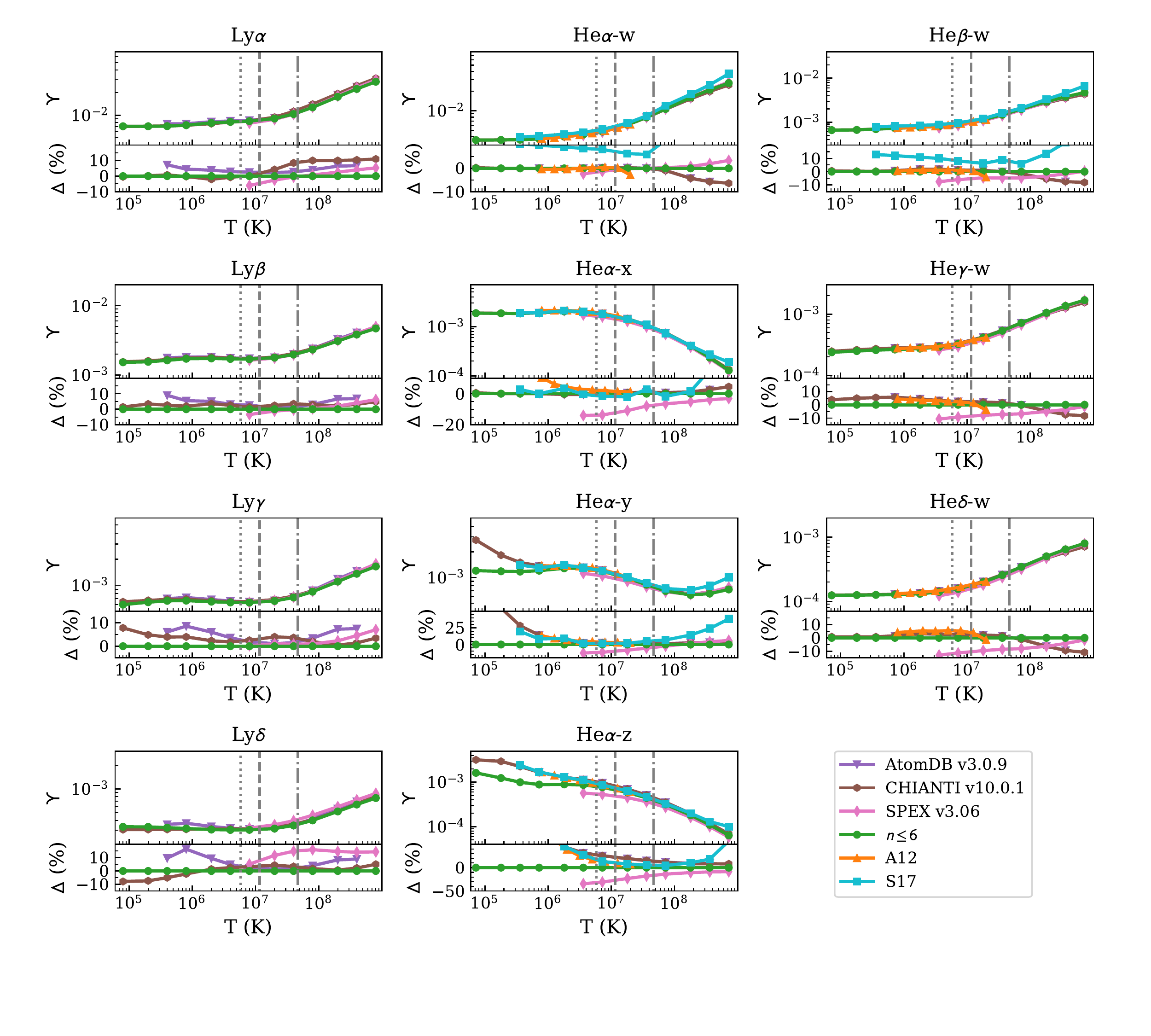}
\caption{Comparison of effective collisional strength data of key diagnostic lines for Ca {\sc xx} and Ca {\sc xix}. AtomDB, CHIANTI, and SPEX data are shown in purple (triangle down), brown (hexagon), and magenta (diamond), respectively. Present work, \citet{Aggarwal2012b}, and \citet{SiR2017} are shown in green (circle), orange (triangle up), and cyan (square), respectively. Percentage difference ($\Delta$) is given with respect to the present work. Vertical dashed lines mark typical temperatures of hot plasmas in individual galaxies ($\sim0.5$~keV, dotted), groups of galaxies ($\sim1$~keV, dashed), and clusters of galaxies ($\sim4$~keV, dot-dashed), respectively. When the ionic fraction is too low (Fig.~\ref{fig:plot_cfcsd}), SPEX skipped the level population calculation (including the effective collisional strength data) for computational efficiency.}
\label{fig:plot_cfecs_z20}
\end{figure*}

As shown in Figure~\ref{fig:plot_cfecs_z20}, the effective collision strengths of the Lyman series agree $\lesssim10$~\% for Ly$\alpha$ to Ly$\delta$ lines between the present work and \citet[used by CHIANTI and SPEX][]{Aggarwal1992-Ca}. Similar good agreement is found between the present work and \citet[][used by AtomDB]{LiS2015} at $T\gtrsim10^7$~K. At lower temperatures, relatively larger differences (up to $\sim20$~\%) can be found.

For the He$\alpha$-w (resonance) line, all $R$-matrix data agree $\lesssim3$~\% at $T<10^{8}$~K except CHIANTI at $T\gtrsim10^8$~K. The original $R$-matrix data from \citet{Aggarwal2012b} is calculated up to $10^{7.4}$~K. The high-temperature extrapolation in CHIANTI might be the issue. IPIRDW data from \citet{SiR2017} is systematically higher by $\gtrsim5$~\%, similar to Fe {\sc xxv} (Sect.~\ref{sct:cfecs_z26}). For the He$\alpha$-x (intercombination) line, the \citet{Sampson1983} data (used by SPEX) stands out at $T\sim10^{7-8}$~K but it is still within $\sim20$~\%. For the He$\alpha$-y (intercombination) line, large differences ($\gtrsim25$~\%) can be found between the present work and \citet[][used by AtomDB and CHIANTI]{Whiteford2005} at $T\lesssim10^6$~K. Even larger differences can be found for the He$\alpha$-z line at $T\lesssim10^6$~K, although the ionic fraction of Ca {\sc xix} at $T\lesssim10^{6.2}$~K is more than three orders of magnitude lower than the peak value under CIE conditions (Figure~\ref{fig:plot_cfcsd}). Again, \citet{Whiteford2005} is larger than the present work by $\sim10$~\% at high temperatures ($T\gtrsim10^8$~K). For He$\beta$ lines, the SPEX He$\beta$-z data is again systematically larger than all other calculations at $T\gtrsim10^7$~K.

The comparison of He$\beta$ to He$\delta$ resonance transitions among different data sets share similar issues found for He$\alpha$-w with one caveat. The He$\gamma$-w and He$\delta$-w data in \citet{SiR2017} are systematically lower by one to three orders of magnitude (beyond the plotting frame of Fig.~\ref{fig:plot_cfecs_z20}) when compared to all the $R$-matrix data.

\subsection{Si {\sc xiv} and Si {\sc xiii}}
\label{sct:cfecs_z14}
\begin{figure*}
\centering
\includegraphics[width=\hsize, trim={0.5cm 2.0cm 0.5cm 0.cm}, clip]{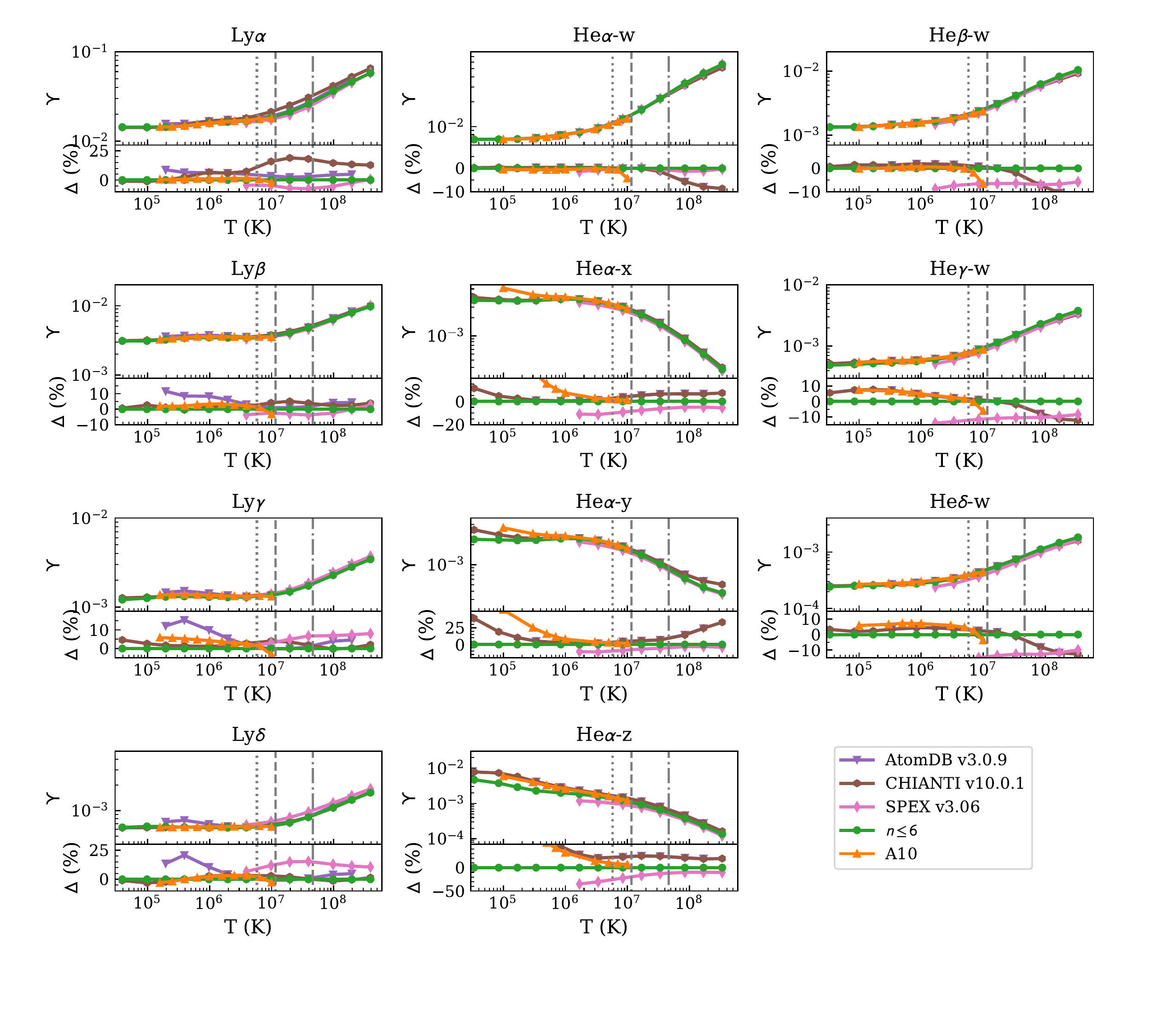}
\caption{Comparison of effective collisional strength data of key diagnostic lines for Si {\sc xiv} and Si {\sc xiii}. AtomDB, CHIANTI, and SPEX data are shown in purple (triangle down), brown (triangle down), and magenta (triangle down), respectively. Present work and \citet{Aggarwal2010b} are shown in green (circle) and orange (triangle up), respectively. Percentage difference ($\Delta$) is given with respect to the present work. Vertical dashed lines mark typical temperatures of hot plasmas in individual galaxies ($\sim0.5$~keV, dotted), groups of galaxies ($\sim1$~keV, dashed), and clusters of galaxies ($\sim4$~keV, dot-dashed), respectively. When the ionic fraction is too low (Fig.~\ref{fig:plot_cfcsd}), SPEX skipped the level population calculation (including the effective collisional strength data) for computational efficiency.}
\label{fig:plot_cfecs_z14}
\end{figure*}

As shown in Figure~\ref{fig:plot_cfecs_z14}, Ly$\alpha$ to Ly$\delta$ agree $\lesssim5$~\% among all $R$-matrix data sets, while the IPIRDW data from \citet{LiS2015} show relatively large (but still within $\sim25$~\%) differences. The high-temperature extrapolation by CHIANTI and SPEX above $10^{6.4}$~K \citep{Aggarwal1992-Si} might explain the difference noticed here.

For He$\alpha$ and He$\delta$ lines, apart from similar issues discussed above, we notice the relatively large ($\gtrsim20$~\%) increase of \citet[][used by CHIANTI]{Whiteford2005} at $T>10^{8}$~K for He$\alpha$-y line. There is no high-temperature extrapolation here because the original calculation goes to $4.5\times10^8$~K. At such a high temperature, the ionic fraction of Si {\sc xiii} is more than three orders of magnitude lower than the peak value under CIE conditions (Figure~\ref{fig:plot_cfcsd}).

\subsection{O {\sc viii} and O {\sc vii}}
\label{sct:cfecs_z08}
The effective collision strengths of the Lyman series agree $\lesssim5$~\% for Ly$\alpha$ and Ly$\beta$ and $\lesssim10$~\% for Ly$\gamma$ and Ly$\delta$ among the present work, AtomDB, CHIANTI, and \citet{Aggarwal2010a}. The original data of \citet{Ballance2003} was calculated up to $1.8\times10^7$~K. At this boundary temperature, the \citet{Ballance2003} data is lower $\lesssim5$~\% than the present work. This smaller difference is likely due to the coverage of scattering energy in the two calculations. \citet{Ballance2003} used 70 continuum basis orbitals to cover at least $\sim4.5$ times the ionization potential of O {\sc viii}. In the present work, we used 110 continuum basis orbitals to cover at least $\sim6.2$ times the ionization potential. As shown in Figure~3 of \citet{Malespin2011}, covering a wider energy range can better constrain the high-temperature effective collision strength. At $T\gtrsim1.8\times10^7$~K, AtomDB and CHIANTI extrapolated the high-temperature data differently. In addition, the SPEX Ly$\beta$ data is systematically lower than the present work by $\gtrsim10$~\% at $T\lesssim10^8$~K.

\begin{figure*}
\centering
\includegraphics[width=\hsize, trim={0.5cm 2.0cm 0.5cm 0.cm}, clip]{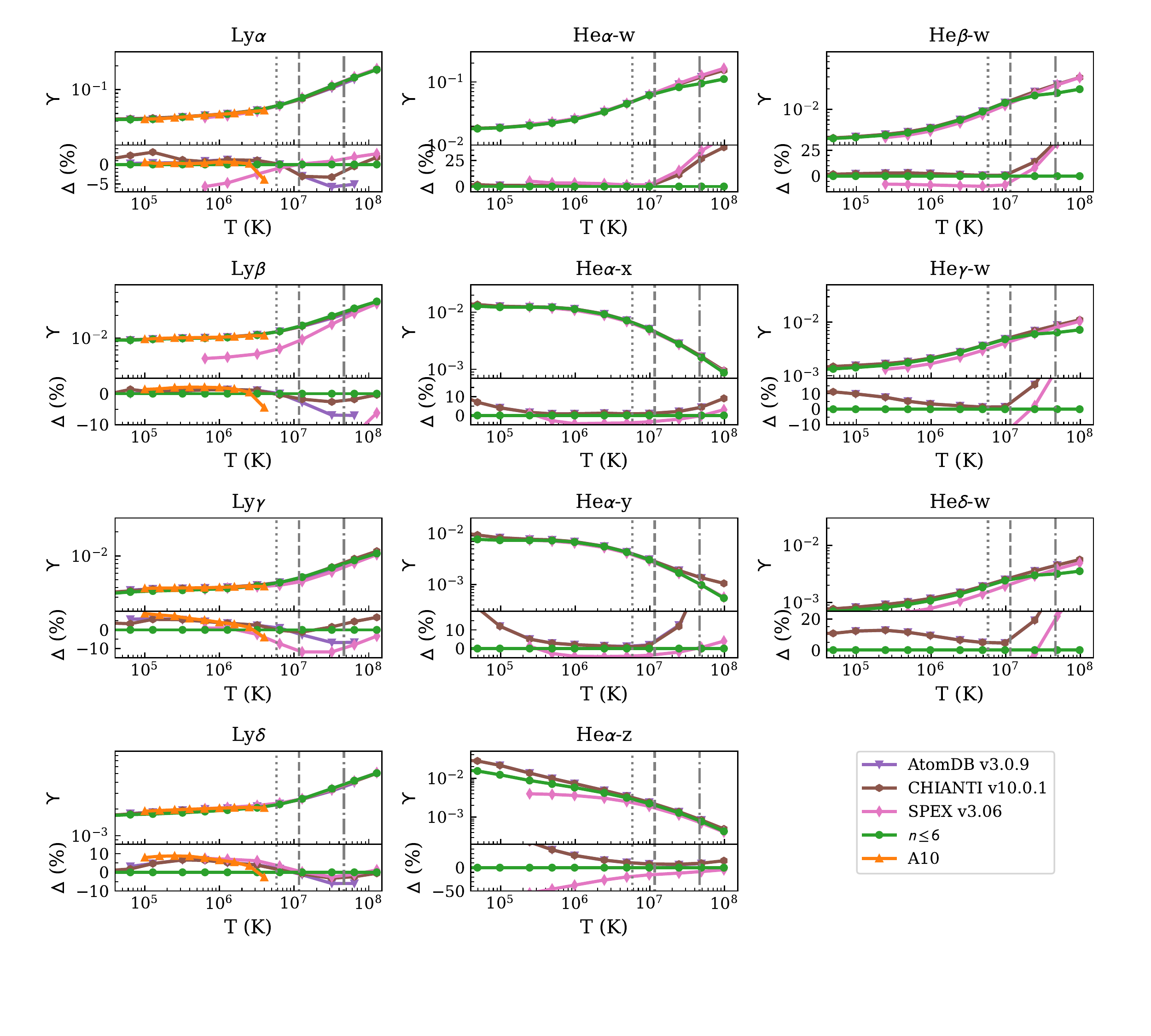}
\caption{Comparison of effective collisional strength data of key diagnostic lines for O {\sc viii} and O {\sc vii}. AtomDB, CHIANTI, and SPEX data are shown in purple (triangle down), brown (triangle down), and magenta (triangle down), respectively. Present work and \citet{Aggarwal2010a} are shown in green (circle) and yellow (triangle up), respectively. Percentage difference ($\Delta$) is given with respect to the present work. Vertical dashed lines mark typical temperatures of hot plasmas in individual galaxies ($\sim0.5$~keV, dotted), groups of galaxies ($\sim1$~keV, dashed), and clusters of galaxies ($\sim4$~keV, dot-dashed), respectively. When the ionic fraction is too low (Fig.~\ref{fig:plot_cfcsd}), SPEX skipped the level population calculation (including the effective collisional strength data) for computational efficiency.}
\label{fig:plot_cfecs_z08}
\end{figure*}

More than $\sim10$~\% differences are found between the present work and \citet{Whiteford2005} for the resonance and intercombination lines at $T\gtrsim4\times10^7$~K. Under CIE conditions, the ionic fraction of O {\sc vii} at $T\gtrsim10^7$~K is more than three orders of magnitude lower than the peak value (Figure~\ref{fig:plot_cfcsd}). For the forbidden line (He$\alpha$-z), $\gtrsim10$~\% difference is noticed at $T<10^6$~K, where the ionic fraction of O {\sc vii} peaks. This temperature is less relevant for studies of individual galaxies and galaxy assemblies, but it might be relevant for stellar coronae.

\subsection{Exemplary impact to observations}
\label{sct:simu}
As mentioned earlier, a relatively large difference ($\gtrsim10$~\% for $T\lesssim10^8$~K) is found between the present work and SPEX v3.06.01 for O {\sc viii} Ly$\beta$, which is less affected by resonance scattering than Ly$\alpha$. Here we show simulated spectra representative of two next-generation high-resolution X-ray spectrometers: HUBS \citep{CuiW2020} and Arcus \citep{Smith2016}. The former employs superconducting transition-edge sensors while the latter adopts critical angle transmission gratings to achieve rather high spectral resolution. The central array of HUBS aims to yield an energy resolution of 0.6 eV in the $0.1-2$~keV energy band. Arcus will achieve $R=\lambda/\Delta \lambda=3800$ in the $10-50$~\AA\ wavelength range when using a very narrow extraction region. Furthermore, the relatively large effective area of both instruments enables observers to obtain high-quality spectra for relatively dim targets (as in the following example).

\begin{figure*}
\centering
\includegraphics[width=.9\hsize, trim={1.5cm 0.5cm 1.0cm 0.cm}, clip]{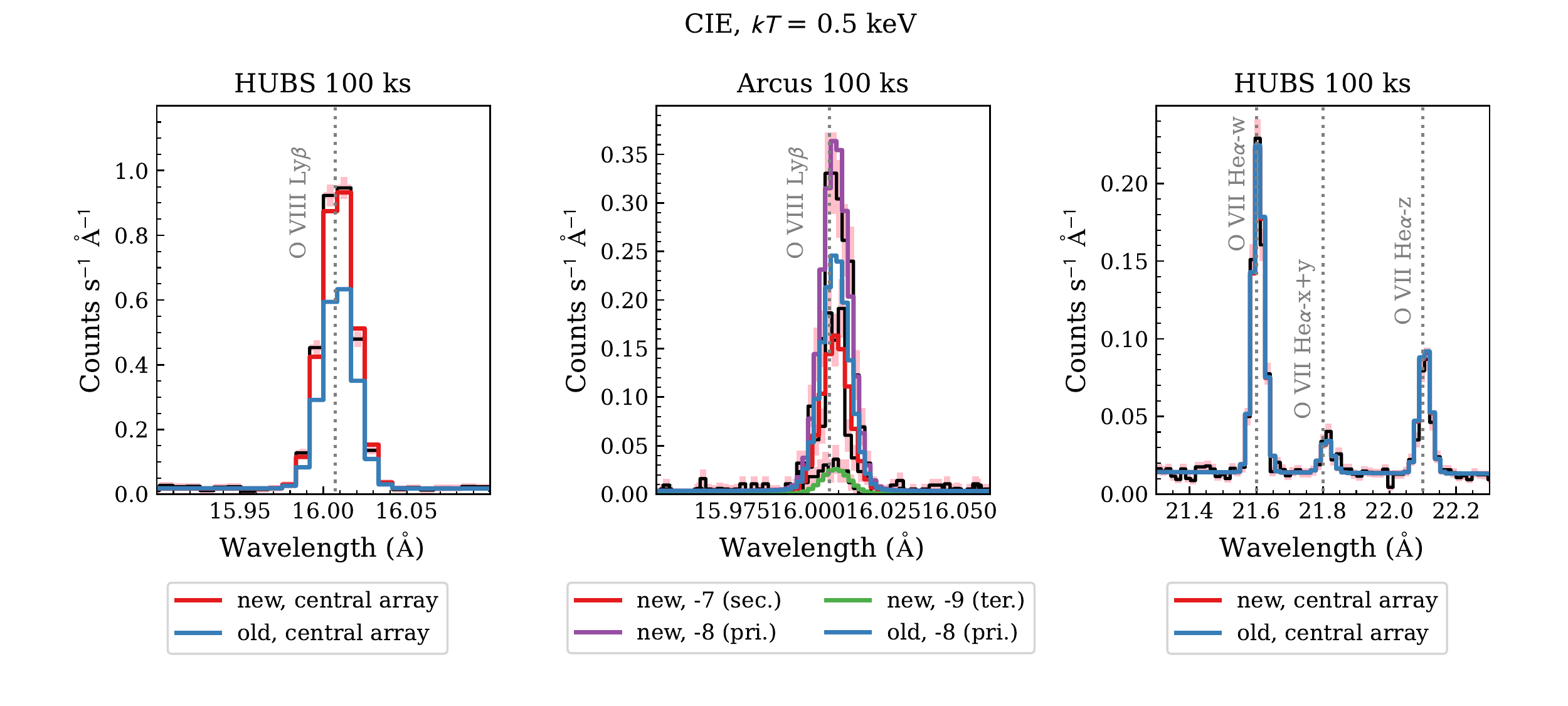}
\caption{Simulated HUBS (left and right) and Arcus (middle) spectra in the O {\sc viii} Ly$\beta$ (left and middle) and O {\sc vii} triplet neighbourhood. Simulated data are shown in black and the $1\sigma$ statistical uncertainties in pink. Model spectra using the new atomic data presented in this work are shown in dashed lines, while those using the atomic data from SPEX v3.06.01 are in solid blue lines. Due to geometry overlapping for Arcus, the O {\sc viii} Ly$\beta$ line comes from the primary spectral order (the $-8$th order, purple) as well as two neighboring spectral orders ($-7$th and $-9$th). See text for the detailed simulation setup (Sect.~\ref{sct:cfecs_z08}). }
\label{fig:plot_cf_dnma_inst}
\end{figure*}

For both instruments, we set an arbitrary exposure time of 100 ks, an observed $0.5-10$~keV flux of $5.0\times10^{-12}~{\rm erg~s^{-1}~cm^{-2}}$, a negligible line-of-sight galactic hydrogen column density of $1.0\times10^{20}~{\rm cm^{-2}}$, a single-temperature CIE plasma with $kT=0.5$~keV. Only the oxygen abundance is set to solar \citep{Lodders2009} while other metal abundances are set to zero. We used SPEX v3.06.01 (the latest released version) for the simulation, as well as a development version where the H- and He-like electron-impact excitation data were updated using the present work. As shown in the left panel of Fig.~\ref{fig:plot_cf_dnma_inst}, the old atomic data would underestimate the O {\sc viii} Ly$\beta$ line flux at the core by $\sim32$~\%, which is a factor of $\sim8$ times larger than the 1 $\sigma$ statistical uncertainty in this HUBS simulation\footnote{We use the HUBS response files (v20201227) of the central array, which has a energy resolution of 0.6 eV.}. In the Arcus simulation\footnote{We use the Arcus response files (6500d8b) with the ``osip60" configuration, which has a better energy resolution but a smaller effective area.}, the O {\sc viii} Ly$\beta$ comes from three geometrical overlapping spectral orders: the $-8$th order (primary), the $-7$th order (secondary), and $-9$th order (tertiary). The flux difference between the old and new atomic data for the primary spectral order is a factor of $\sim6$ larger than the flux of the tertiary spectral order. For the O {\sc vii} triplet, as shown in the right panel of Fig.~\ref{fig:plot_cf_dnma_inst}, the line flux between the old and new atomic data are negligible. The old and new atomic atomic data agree well at $kT=0.5$~keV for the resonance and inter-combination lines (Fig.~\ref{fig:plot_cfecs_z08}), thus we do not expect noticeable differences in the simulated spectra. The old and new atomic data of the forbidden line differ by $\sim20$~\% at $kT=0.5$~keV (Fig.~\ref{fig:plot_cfecs_z08}). But its impact is limited because cascading from upper levels contributes most to the level population (Fig.~\ref{fig:plot_cflpop_z14}).

\section{Summary}
We have presented systematic radiation damped $R$-matrix intermediate-coupling frame transformation calculations of electron-impact excitation data of H- and He-like ions with atomic number $Z=6-30$. For each ion, fine-structure energy levels up to $n=6$ (36 levels for H-like ions and 71 levels for He-like ions) were included in the target configuration interaction and close-coupling collision expansion. Level-resolved effective collision strengths were obtained among these levels over four orders of magnitude in temperature. When compared with existing $R$-matrix or distorted wave data in the atomic databases and literature, generally speaking, relatively good agreements can be found near the peak temperatures of charge state distribution under collisional ionized equilibrium conditions. The new data calculated here are relevant for current and future high-resolution X-ray spectrometers such as the upcoming XRISM in 2023, Athena/X-IFU, HUBS, Arcus, and so on around 2030s.




\begin{acknowledgments}
JM acknowledge useful discussions with Adam Foster, Connor Ballance, Martin O Mullane, Zhu Liu, Guiyun Liang, Mingyue Hao, Ran Si, and Jelle Kaastra. This work is supported by STFC (UK) through the University of Strathclyde UK APAP network grants ST/T000481/1 and ST/V000863/1.
\end{acknowledgments}

%

\vspace{5mm}





\appendix
\section{Level energies and $A$-values of H- and He-like key diagnostic lines}
\label{sct:cflev_cftran}
Accurate level energies are essential to obtain the correct rest frame line energy (or wavelength). The level energies of the upper levels of the key diagnostics lines in Table~\ref{tbl:line_list} agree well among the three databases with only a few exceptions (Tables~\ref{tbl:cflev_hlike} and ~\ref{tbl:cflev_helike}). For H-like ones, the largest difference comes from the level energy of Ar {\sc xviii} ${\rm 2p}~(^2P_{1/2})$ in CHIANTI \citep[taken from][]{Phillips2003}, which is lower than AtomDB/SPEX by 0.4 eV. For He-like ones, the largest difference comes from the level energy of Mg {\sc xi} ${\rm 1s~3p~^1P_1}$ in SPEX, which is lower by 0.7 eV than AtomDB/CHIANTI. This can be comparable to the energy gain correction of the instrument, as shown in the analysis of the Hitomi/SXS spectrum of the Perseus galaxy cluster \citep{Hitomi-C2018V}. The derived bulk velocity of the intracluster media differs by $6~{\rm km~s^{-1}}$ between SPEX v3.03 and AtomDB v3.0.8, while the energy gain correction of the instrument is $14~{\rm km~s^{-1}}$.

As discussed in Sect.~\ref{sct:dis}, the AS-RM energy levels are less accurate than those that can be obtained from AUTOSTRUCTURE without the restrictions imposed by their use by the Breit--Pauli $R$-matrix code. The latter is denoted as AS-REL. The AS-REL energy levels are also shown in Tables~\ref{tbl:cflev_hlike} and \ref{tbl:cflev_helike}. For H-like ions, the inclusion by AUTOSTRUCTURE of the quantum electrodynamic (QED) effects (vacuum polarization and electron self-energy) reduce the inaccuracy from $\lesssim0.05$~\% to $\lesssim0.001$~\% when compared with the three atomic databases. For He-like ions: at low-charge the two-body Coulomb interaction is the main source of uncertainty ($\lesssim0.23$~\% when compared with the three atomic databases); while at high-charge relativistic effects are the main source and the inclusion as well by AUTOSTRUCTURE of the two-body relativistic interactions reduces the overall inaccuracy (from $\lesssim0.15$~\% to $\lesssim0.03$~\% for Ni XXVII when compared with the three atomic databases).

As shown in Table~\ref{tbl:cftran_hlike}, for the $np~^2P_{3/2,1/2},~n=2-5$ energy levels in H-like ions, the $A$-values in AtomDB, CHIANTI, and SPEX agree well ($\lesssim5$\%) for the resonance lines. Larger deviations (up to 40\%) can be found for energy levels in He-like ions (Table~\ref{tbl:cftran_helike}), especially the upper levels of some intercombination lines and forbidden lines.

The A-values shown in Tables ~\ref{tbl:cftran_hlike} and \ref{tbl:cftran_helike} can differ by up to $\sim40$~\%, depending on which databases are compared. For instance, the A-value of the C {\sc v} He$\alpha$-y line differs between AS-RM and SPEX by $\sim40$~\%, while the AS-RM and AtomDB values are identical. The A-value for this transition in CHIANTI is $\sim27$~\% larger than the AS-RM/AtomDB ones. We are limited to using non-relativistic orbitals (by our scattering calculation) but the perturbative one-body relativistic operators (mass-velocity and Darwin) become increasingly large as the ion charge increases and imbalance the level mixing which in turn leads to the relatively low accuracy for the AS-RM He$\gamma$-w, He$\delta$-w transition rates of high-$Z$ elements. Using (kappa-averaged) relativistic orbitals in AUTOSTRUCTURE eliminates the perturbative imbalance. This can be seen most clearly for H-like ions (where the databases agree well.) The AS-RM Ni {\sc xxviii} Ly$\delta$ A-values in Table~\ref{tbl:cftran_hlike} are $\sim$20\% smaller than the database ones. AUTOSTRUCTURE calculations using relativistic orbitals reduce the difference to a few percent, depending on the database. These transitions are shown in the last columns of Tables~\ref{tbl:cftran_hlike} and \ref{tbl:cftran_helike}. The corresponding Ni {\sc xxvii} He$\gamma$-w, He$\delta$-w transition rates now agree with CHIANTI to within 3\%. Both the AS-RM and AS-REL data sets (in the \textit{adf04} format) are available in the Zenodo package.

\bibliographystyle{aasjournal}
\bibliography{refs}



\startlongtable
\begin{deluxetable*}{l|cccc|cc}
\centering
\footnotesize
\tablecaption{Energy levels (in eV) of H-like ions in SPEX v3.06.01, AtomDB v3.0.9, and CHIANTI database v10.0.1.
The numbers in the parentheses are the differences (in eV) with respect to SPEX of the other two databases. The penultimate column shows the results of the present AUTOSTRUCTURE calculations which were used for the R-matrix calculations (AS-RM). The last column shows the results of AUTOSTRUCTURE calculations including relativistic effects (AS-REL) that are necessarily omitted by AS-RM. See the discussion in Sect.~\ref{sct:dis} and Appendix~\ref{sct:cflev_cftran}.
\label{tbl:cflev_hlike}}
\tablehead{
\colhead{Ion} & \colhead{Level} & \colhead{SPEX} &
\colhead{AtomDB} & \colhead{CHIANTI} & \colhead{AS-RM} & \colhead{AS-REL}
}
\startdata
C {\sc vi} & ${\rm 2p~^2P_{1/2}}$ & 367.5 & 367.5 (-0.0) & 367.5 (-0.0) & 367.5 & 367.5 \\
C {\sc vi} & ${\rm 2p~^2P_{3/2}}$ & 367.5 & 367.5 (-0.0) & 367.5 (-0.0) & 367.6 & 367.6 \\
C {\sc vi} & ${\rm 3p~^2P_{1/2}}$ & 435.5 & 435.5 (-0.0) & 435.5 (-0.0) & 435.6 & 435.6 \\
C {\sc vi} & ${\rm 3p~^2P_{3/2}}$ & 435.6 & 435.6 (-0.0) & 435.6 (-0.0) & 435.6 & 435.6 \\
C {\sc vi} & ${\rm 4p~^2P_{1/2}}$ & 459.4 & 459.4 (-0.0) & 459.4 (-0.0) & 459.4 & 459.4 \\
C {\sc vi} & ${\rm 4p~^2P_{3/2}}$ & 459.4 & 459.4 (-0.0) & 459.4 (-0.0) & 459.4 & 459.4 \\
C {\sc vi} & ${\rm 5p~^2P_{1/2}}$ & 470.4 & 470.4 (-0.0) & 470.4 (-0.0) & 470.4 & 470.4 \\
C {\sc vi} & ${\rm 5p~^2P_{3/2}}$ & 470.4 & 470.4 (-0.0) & 470.4 (-0.0) & 470.4 & 470.4 \\
\enddata
\end{deluxetable*}

\startlongtable
\begin{deluxetable*}{l|cccc|cc}
\centering
\footnotesize
\tablecaption{Energy levels (in eV) of He-like ions in SPEX v3.06.01, AtomDB v3.0.9, and CHIANTI database v10.0.1.
The numbers in the parentheses are the differences (in eV) with respect to SPEX of the other two databases. The penultimate column shows the results of the present AUTOSTRUCTURE calculations which were used for the R-matrix calculations (AS-RM). The last column shows the results of AUTOSTRUCTURE calculations including relativistic effects (AS-REL) that are necessarily omitted by AS-RM. See the discussion in Sect.~\ref{sct:dis} and Appendix~\ref{sct:cflev_cftran}.
\label{tbl:cflev_helike}}
\tablehead{
\colhead{Ion} & \colhead{Level} & \colhead{SPEX} &
\colhead{AtomDB} & \colhead{CHIANTI} & \colhead{AS-RM} & \colhead{AS-REL}
}
\startdata
C {\sc v} & ${\rm 1s~2p~^1P_{1}}$ & 307.9 & 307.9 (+0.0) & 307.9 (+0.0) & 308.6 & 308.5 \\
C {\sc v} & ${\rm 1s~2p~^3P_{2}}$ & 304.4 & 304.4 (+0.0) & 304.4 (+0.0) & 304.5 & 304.5 \\
C {\sc v} & ${\rm 1s~2p~^3P_{1}}$ & 304.4 & 304.4 (+0.0) & 304.4 (+0.0) & 304.5 & 304.4 \\
C {\sc v} & ${\rm 1s~2s~^3S_{1}}$ & 299.0 & 299.0 (+0.0) & 299.0 (+0.0) & 299.0 & 298.9 \\
C {\sc v} & ${\rm 1s~3p~^1P_{1}}$ & 354.5 & 354.5 (-0.0) & 354.5 (-0.0) & 354.9 & 354.8 \\
C {\sc v} & ${\rm 1s~4p~^1P_{1}}$ & 370.9 & 370.9 (-0.0) & 370.9 (-0.0) & 371.3 & 371.2 \\
C {\sc v} & ${\rm 1s~5p~^1P_{1}}$ & 378.5 & 378.5 (-0.0) & 378.5 (-0.0) & 378.9 & 378.8 \\
\noalign{\smallskip}
N {\sc vi} & ${\rm 1s~2p~^1P_{1}}$ & 430.7 & 430.7 (-0.0) & 430.7 (+0.0) & 431.4 & 431.3 \\
\enddata
\end{deluxetable*}

\startlongtable
\begin{deluxetable*}{cl|lllll|ll}
\centering
\footnotesize
\tablecaption{Lyman transition data of H-like ions in SPEX v3.06.01, AtomDB v3.0.9, CHIANTI database v10.0.1, and the present work (AUTOSTRUCTURE)}. The wavelengths (as in SPEX) are in units of \AA\ and $A$-values in ${\rm s^{-1}}$. The column labelled $\Delta$ is the maximum percentage deviation ((max - min) / max $\times100$\%) amongst the three databases.  The penultimate column shows the results of the present AUTOSTRUCTURE calculations which were used for the R-matrix calculations (AS-RM). The last column shows the results of AUTOSTRUCTURE calculations including relativistic effects (AS-REL) that are necessarily omitted by AS-RM. See the discussion in Sect.~\ref{sct:dis} and Appendix~\ref{sct:cflev_cftran}.
\label{tbl:cftran_hlike}
\tablehead{
\colhead{Ion} & \colhead{ID} & \colhead{$\lambda$~(\AA)} & \colhead{SPEX} &
\colhead{AtomDB} & \colhead{CHIANTI} & $\Delta$ & \colhead{AS-RM} & \colhead{AS-REL}
}
\startdata
C {\sc vi} & Ly$\alpha_{1/2}$ &33.740 & $8.12\times10^{11}$ & $8.12\times10^{11}$ & $8.12\times10^{11}$ & 0.1\% & $8.12\times10^{11}$ & $8.12\times10^{11}$ \\
C {\sc vi} & Ly$\alpha_{3/2}$ &33.734 & $8.12\times10^{11}$ & $8.12\times10^{11}$ & $8.12\times10^{11}$ & 0.1\% & $8.12\times10^{11}$ & $8.12\times10^{11}$ \\
C {\sc vi} & Ly$\beta_{1/2}$ &28.466 & $2.17\times10^{11}$ & $2.17\times10^{11}$ & $2.16\times10^{11}$ & 0.3\% & $2.16\times10^{11}$ & $2.17\times10^{11}$ \\
C {\sc vi} & Ly$\beta_{3/2}$ &28.465 & $2.17\times10^{11}$ & $2.17\times10^{11}$ & $2.16\times10^{11}$ & 0.3\% & $2.16\times10^{11}$ & $2.17\times10^{11}$ \\
C {\sc vi} & Ly$\gamma_{1/2}$ &26.990 & $8.84\times10^{10}$ & $8.83\times10^{10}$ & $8.77\times10^{10}$ & 0.8\% & $8.78\times10^{10}$ & $8.82\times10^{10}$ \\
C {\sc vi} & Ly$\gamma_{3/2}$ &26.990 & $8.84\times10^{10}$ & $8.84\times10^{10}$ & $8.79\times10^{10}$ & 0.5\% & $8.80\times10^{10}$ & $8.85\times10^{10}$ \\
C {\sc vi} & Ly$\delta_{1/2}$ &26.357 & $4.46\times10^{10}$ & $4.45\times10^{10}$ & $4.38\times10^{10}$ & 1.7\% & $4.41\times10^{10}$ & $4.45\times10^{10}$ \\
C {\sc vi} & Ly$\delta_{3/2}$ &26.357 & $4.46\times10^{10}$ & $4.46\times10^{10}$ & $4.40\times10^{10}$ & 1.4\% & $4.42\times10^{10}$ & $4.46\times10^{10}$ \\
\enddata
\end{deluxetable*}

\startlongtable
\begin{deluxetable*}{cl|lllll|ll}
\centering
\footnotesize
\tablecaption{Transition data of He-like ions in SPEX v3.06.01, AtomDB v3.0.9, and CHIANTI database v10.0.1, and the present work (AUTOSTRUCTURE)}. The wavelengths (as in SPEX) are in units of \AA\ and $A$-values in ${\rm s^{-1}}$. The column labelled $\Delta$ is the maximum percentage deviation ((max - min) / max $\times100$\%) amongst the three databases.  The penultimate column shows the results of the present AUTOSTRUCTURE calculations which were used for the R-matrix calculations (AS-RM). The last column shows the results of AUTOSTRUCTURE calculations including relativistic effects (AS-REL) that are necessarily omitted by AS-RM. See the discussion in Sect.~\ref{sct:dis} and Appendix~\ref{sct:cflev_cftran}.
\label{tbl:cftran_helike}
\tablehead{
\colhead{Ion} & \colhead{ID} & \colhead{$\lambda$~(\AA)} & \colhead{SPEX} &
\colhead{AtomDB} & \colhead{CHIANTI} & $\Delta$ & \colhead{AS-RM} & \colhead{AS-REL}
}
\startdata
C {\sc v} & He$\alpha$-z & 41.472 & $4.96\times10^{1}$ & $4.05\times10^{1}$ & $4.05\times10^{1}$ & 18.3\% & $4.05\times10^{1}$ & $4.53\times10^{1}$ \\
C {\sc v} & He$\alpha$-y & 40.731 & $2.89\times10^{7}$ & $1.76\times10^{7}$ & $2.24\times10^{7}$ & 39.1\% & $1.76\times10^{7}$ & $2.14\times10^{7}$ \\
C {\sc v} & He$\alpha$-x & 40.728 & $2.65\times10^{4}$ & $2.49\times10^{4}$ & $2.54\times10^{4}$ & 6.0\% & $2.53\times10^{4}$ & $2.52\times10^{4}$ \\
C {\sc v} & He$\alpha$-w & 40.268 & $8.87\times10^{11}$ & $9.34\times10^{11}$ & $9.23\times10^{11}$ & 5.0\% & $9.33\times10^{11}$ & $9.33\times10^{11}$ \\
C {\sc v} & He$\beta$-w & 34.973 & $2.55\times10^{11}$ & $2.84\times10^{11}$ & $2.78\times10^{11}$ & 10.1\% & $2.82\times10^{11}$ & $2.83\times10^{11}$ \\
C {\sc v} & He$\gamma$-w & 33.426 & $1.07\times10^{11}$ & $1.25\times10^{11}$ & $1.21\times10^{11}$ & 14.8\% & $1.23\times10^{11}$ & $1.24\times10^{11}$ \\
C {\sc v} & He$\delta$-w & 32.754 & $5.43\times10^{10}$ & $6.97\times10^{10}$ & $6.42\times10^{10}$ & 22.2\% & $6.60\times10^{10}$ & $6.66\times10^{10}$ \\
N {\sc vi} & He$\alpha$-z & 29.534 & $2.57\times10^{2}$ & $2.17\times10^{2}$ & $2.17\times10^{2}$ & 15.6\% & $2.17\times10^{2}$ & $2.39\times10^{2}$ \\
\enddata
\end{deluxetable*}

\end{document}